\documentclass[twocolumn,showpacs,preprintnumbers,amsmath,amssymb]{revtex4}
\usepackage{graphicx}
\usepackage{dcolumn}
\usepackage{bm}

\def\be{\begin{equation}}
\def\ee{\end{equation}}
\def\ba{\begin{array}}
\def\ea{\end{array}}
\def\bea{\begin{eqnarray}}
\def\eea{\end{eqnarray}}

\begin{document}
\preprint{APS/123-QED}
\title{\large\bf Role of colliding geometry on the balance energy of mass asymmetric systems.}

\author{Supriya Goyal}%
\email{ashuphysics@gmail.com}

\affiliation{\it Department of Physics, Panjab University, \\
Chandigarh -160 014, India\\ }


\date{\today}

\begin{abstract}
We study the role of colliding geometry on the balance energy
($E_{bal}$) of mass-asymmetric systems by varying the mass
asymmetry ($\eta$ = $A_{T}-A_{P}$/$A_{T}+A_{P}$; where $A_{T}$ and
$A_{P}$ are the masses of the target and projectile, respectively)
from 0.1 to 0.7, over the mass range 40-240 and on the mass
dependence of the balance energy. Our findings reveal that
colliding geometry has a significant effect on the $E_{bal}$ of
asymmetric systems. We find that, as we go from central collisions
to peripheral ones, the effect of mass asymmetry on $E_{bal}$
increases throughout the mass range. Interestingly, we find that
for every fixed system mass ($A_{TOT}$), the effect of the impact
parameter variation is almost uniform throughout the
mass-asymmetry range. For each $\eta$, $E_{bal}$ follows a
power-law behavior ($\propto A^{\tau}$) at all colliding
geometries.
\end{abstract}

\maketitle

In the intermediate energy region (i.e., from 10 MeV/nucleon to 1
GeV/nucleon), there are two major interactions that dominate the
nuclear reaction: (i) the attractive nuclear mean field and (ii)
the repulsive nucleon-nucleon (nn) interaction. Among the various
phenomena observed in this energy region, collective transverse
flow is one of the most sensitive and sought-after and has been
extensively used over the past three decades to provide
information about the nuclear equation of state (EOS) and nn cross
section \cite{1,2,3}. The collective transverse flow is the
sideward deflection of the reaction products in phase space and is
due to the interactions inside the reaction zone. At low beam
energies, the collective transverse flow is dominated by an
attractive interaction and the flow is expected to be negative. At
high beam energies, the flow is dominated by nn repulsive
interactions and the flow is expected to be positive. At a beam
energy where the repulsive nn interactions balance the attractive
nuclear mean-field interactions, the collective transverse flow
disappears. The beam energy at which transverse flow disappears is
termed the {\it balance energy} ($E_{bal}$) \cite{4}.

The balance energy is found to be highly sensitive toward the
nuclear matter equation of state, the nn cross section
\cite{1,2,3}, the size of the system \cite{5}, the asymmetry of
the reaction \cite{6}, the incident energy of the projectile
\cite{7}, and the colliding geometry (i.e., the impact parameter)
\cite{8}.

The importance of the role of system mass \cite{9a,9,10} and
colliding geometry \cite{11,11a,12,12a} in the determination of
the $E_{bal}$ has been recognized several times in the literature.
For the central collisions, the balance energy is found to vary as
$A_{TOT}^{-1/3}$ (where $A_{TOT}$ is the total mass of the target
and projectile), both experimentally as well as theoretically
\cite{9a}. Owing to the decrease in the compression reached in
heavy-ion collisions with increase in the impact parameter,
$E_{bal}$ is found to increase approximately linearly as a
function of impact parameter \cite{11a,12}. It is noted that the
role of the impact parameter on the disappearance of flow and on
its mass dependence has been studied mainly for symmetric reacting
nuclei. However, in a recent study \cite{6}, we confronted for the
first time the effect of mass asymmetry ($\eta$) on $E_{bal}$
using a quantum molecular dynamics (QMD) model \cite{13}, but the
study was limited to central collisions only ($b/b_{max}$ = 0.25).
We found that for central collisions, almost independent of the
system mass, a uniform effect of mass asymmetry can be seen at the
balance energy. Here we aim to extend the study by looking for the
effect of colliding geometries on the $E_{bal}$ of mass-asymmetric
systems by varying the impact parameter from central to peripheral
values. This study is performed with the QMD model explained in
Ref. \cite{13}.
\begin{figure}[!t]
\centering \vskip - 1.0cm
\includegraphics* [scale=0.40] {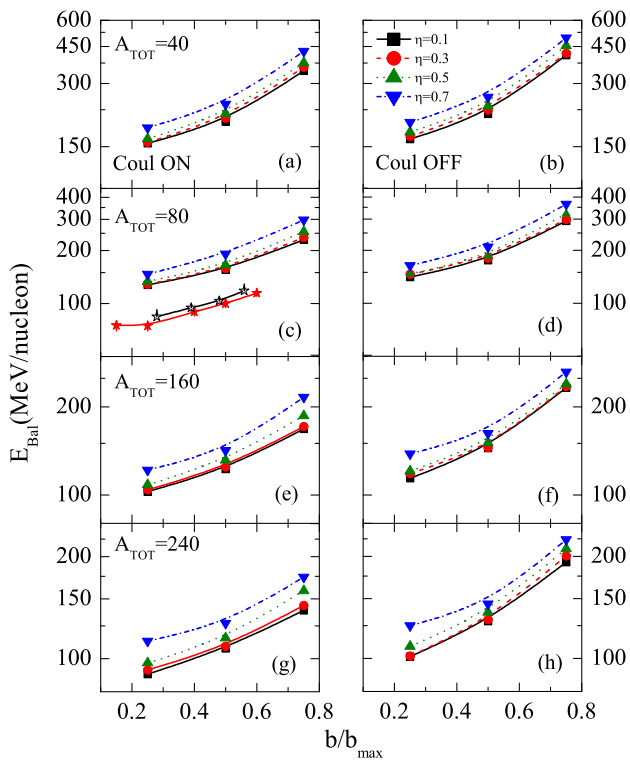}
\vskip -0.7cm \caption {(Color online) $E_{bal}$ as a function of
reduced impact parameter (a, c, e, g) with and (b, d, f, h)
without Coulomb potential for different system masses. The results
for different asymmetries $\eta$ = 0.1, 0.3, 0.5, and 0.7 are
represented, respectively, by the solid squares, circles,
triangles, and inverted triangles. Solid (open) stars in (c)
represent data points of $^{40}Ca+^{40}Ca$ ($^{40}Ar+^{45}Sc$) and
are taken from Ref. \cite{12}. Lines are a guide for the eye.}
\end{figure}
\begin{table*}
\caption{The mean values of percentage change in $E_{bal}$ with
impact parameter for $A_{TOT}$ = 40-240 with and without Coulomb
potential.}
\begin{center}
\begin{tabular}{|c|c|c|c|c|}
\hline System
&\multicolumn{2}{c|}{Coul. ON}&\multicolumn{2}{c|}{Coul.
OFF}\\\cline{2-5}

 &$b/b_{max}=0.5$ &$b/b_{max}=0.75$ &$b/b_{max}=0.5$ &$b/b_{max}=0.75$\\\cline{2-5}
 \hline 40   &29.80  &127.69 &32.43 &152.04 \\\cline{2-5}

 \hline 80   &24.92  &90.07 &25.22 &113.91 \\\cline{2-5}

 \hline 160  &19.28  &69.38 &22.95 &97.66 \\\cline{2-5}

 \hline 240  &16.88  &56.47 &24.28 &89.98 \\\cline{2-5}

 \hline
\end{tabular}
\end{center}
\end{table*}
\begin{figure}[!t]
\centering \vskip -1.0cm
\includegraphics* [scale=0.40] {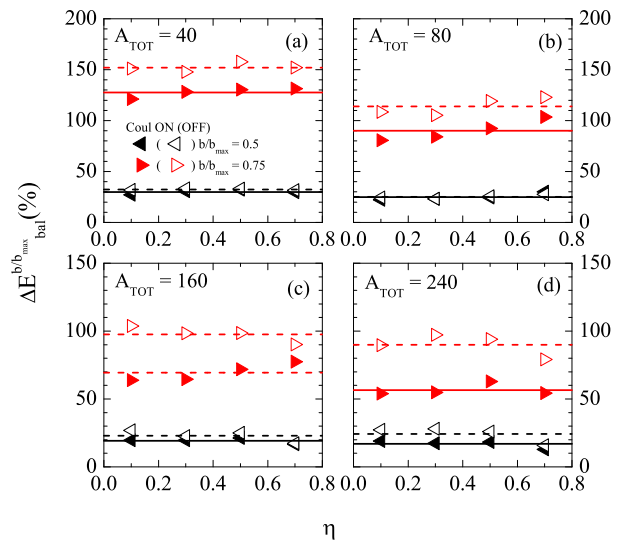}
\vskip -2.5cm \caption {(Color online) The percentage difference
$\Delta E_{bal}^{b/b_{max}}$(\%) as a function of ${\eta}$ for
different system masses. The results of the percentage difference
for different colliding geometries $b/b_{max}$ = 0.5 and 0.75 are
represented, respectively, by the left- and right-pointing
triangles. Open symbols represents results without Coulomb
potential. Horizontal dotted lines represent the mean value of
$\Delta E_{bal}^{b/b_{max}}$(\%) for each
$b/b_{max}$.}\label{fig2}
\end{figure}
\begin{figure}[!t]
\centering \vskip -1.0 cm
\includegraphics* [scale=0.40] {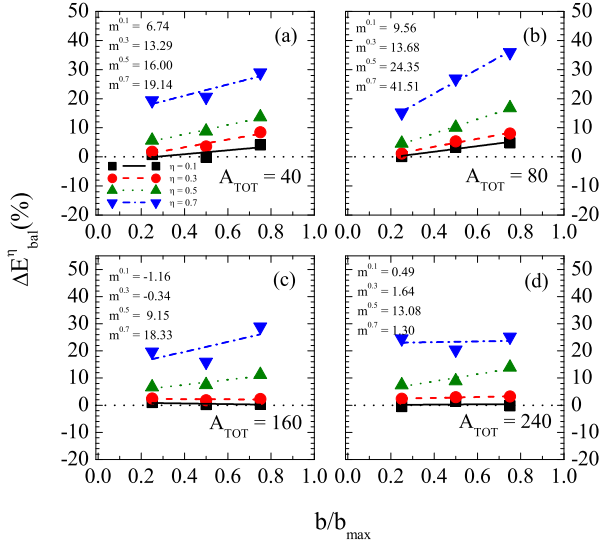}
\vskip -2.5cm \caption {(Color online) The percentage difference
$\Delta E_{bal}^{\eta}$(\%) as a function of $b/b_{max}$ for
different system masses. The results of the percentage difference
for different asymmetries $\eta$ = 0.1, 0.3, 0.5, and 0.7 are
represented, respectively, by the solid squares, circles,
triangles, and inverted triangles. Lines are the linear fits
($\propto m\frac {b}{b_{max}}$); {\it m} values without errors are
displayed. Results are with Coulomb potential.}\label{fig3}
\end{figure}
\begin{figure}[!t]
\centering \vskip -1.0 cm
\includegraphics* [scale=0.40]{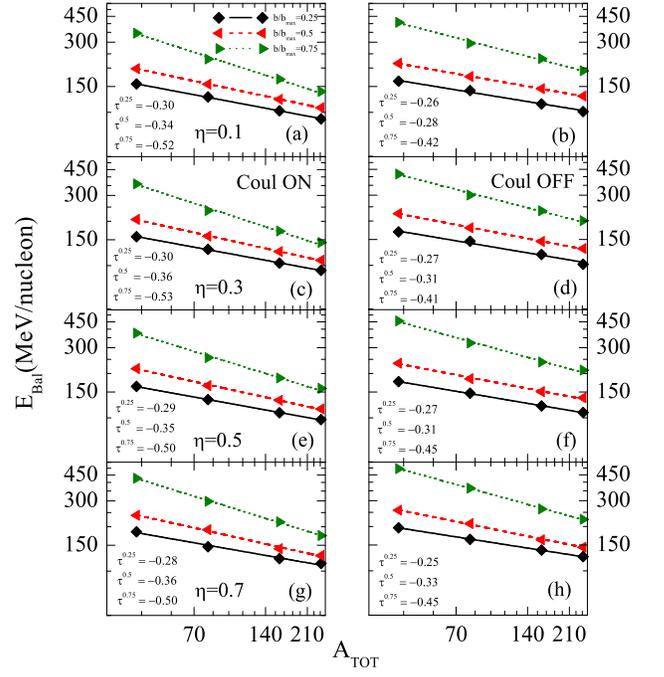}
\vskip -0.7cm \caption {(Color online) Left (right) panels display
$E_{bal}$ as a function of the combined mass of the system for
different mass asymmetries (a, c, e, g) with and (b, d, f, h)
without Coulomb potential. Solid diamonds, left-pointing
triangles, and right-pointing triangles represents the
calculations with $b/b_{max}$ = 0.25, 0.5, and 0.75, respectively.
Lines are power-law fits $\propto A^{\tau}_{TOT}$; ${\tau}$ values
without errors are displayed.}\label{fig4}
\end{figure}
\begin{figure}[!t]
\centering
\vskip -1.0cm
\includegraphics* [scale=0.40] {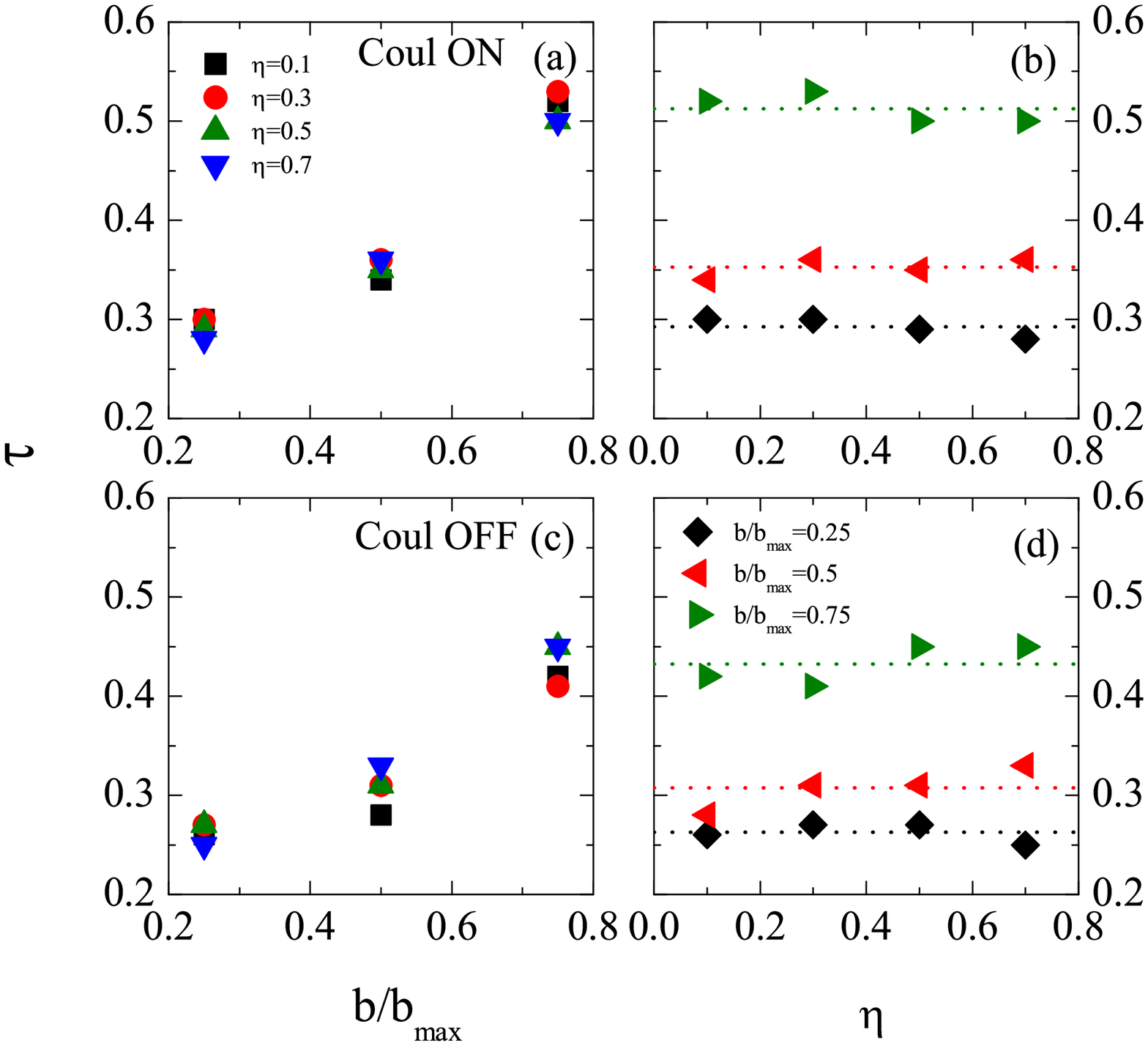}
\vskip -2.5cm \caption {(Color online) Values of ${\tau}$ as a
function of (a, c) reduced impact parameter for ${\eta}$ = 0.1-0.7
and (b, d) mass asymmetry for $b/b_{max}$ = 0.25-0.75. ${\tau}$
values (a, b) with (c, d) without Coulomb potential. The
horizontal dotted lines in (b) and (d) represent the mean value of
${\tau}$ for each $b/b_{max}$. Various symbols have the same
meaning as in Figs. 1 and 4.}\label{fig5}
\end{figure}

For the present study, we simulated 1000-5000 events of various
reactions in the incident energy range between 90 and 600
MeV/nucleon. In particular, we simulated the reactions of
$^{17}_{8}O+^{23}_{11}Na$ ($\eta = 0.1$),
$^{14}_{7}N+^{26}_{12}Mg$ ($\eta = 0.3$),
$^{10}_{5}B+^{30}_{14}Si$ ($\eta = 0.5$), and
$^{6}_{3}Li+^{34}_{16}S$ ($\eta = 0.7$) for $A_{TOT}$ = 40;
$^{36}_{18}Ar+^{44}_{20}Ca$ ($\eta = 0.1$),
$^{28}_{14}Si+^{52}_{24}Cr$ ($\eta = 0.3$),
$^{20}_{10}Ne+^{60}_{28}Ni$ ($\eta = 0.5$), and
$^{10}_{5}B+^{70}_{32}Ge$ ($\eta = 0.7$) for $A_{TOT}$ = 80;
$^{70}_{32}Ge+^{90}_{40}Zr$ ($\eta = 0.1$),
$^{54}_{26}Fe+^{106}_{48}Cd$ ($\eta = 0.3$),
$^{40}_{20}Ca+^{120}_{52}Te$ ($\eta = 0.5$), and
$^{24}_{12}Mg+^{136}_{58}Ce$ ($\eta = 0.7$) for $A_{TOT}$ = 160;
and $^{108}_{48}Cd+^{132}_{56}Ba$ ($\eta = 0.1$),
$^{84}_{38}Sr+^{156}_{66}Dy$ ($\eta = 0.3$),
$^{60}_{28}Ni+^{180}_{74}W$ ($\eta = 0.5$), and
$^{36}_{18}Ar+^{204}_{82}Pb$ ($\eta = 0.7$) for $A_{TOT}$ = 240.
The present study is performed for three values of reduced impact
parameter (i.e., b/b$_{max}$ = 0.25, 0.5, and 0.75). The charges
are chosen in such a way that colliding nuclei are stable
nuclides; therefore, slight variation in the charges can be seen.
Here $\eta$ is varied from 0.1 to 0.7, keeping $A_{TOT}$ fixed.
The values of $A_{TOT}$ vary from 40 to 240. A momentum-dependent
soft equation of state with standard energy-dependent Cugnon cross
section (labeled SMD) is used in the present reactions. The value
of $E_{bal}$ is calculated using the {\it directed transverse
momentum $<P^{dir}_{x}>$} and is defined in Ref. \cite{6}.

In Fig. 1, we display the impact parameter dependence of $E_{bal}$
for different values of $\eta$ = 0.1, 0.3, 0.5, and 0.7. The value
of $A_{TOT}$ is kept fixed as 40 [Figs 1(a) and 1(b)], 80 [Figs
1(c) and 1(d)], 160 [Figs 1(e) and 1(f)], and 240 [Figs 1(g) and
1(h)]. Various symbols are explained in the caption of the figure.
From the figure it is clear that, for all mass ranges and
colliding geometries, the value of $E_{bal}$ increases with
increase in $\eta$. This is because, with increase in $\eta$, the
maximum density and compression reached in the overlap zone
decrease, which leads to the decrease in nn collisions.
Furthermore, for all the mass ranges and $\eta$, the well-known
trend of the increase of $E_{bal}$ with impact parameter is seen.
To precisely see the effect of mass asymmetry at different
colliding geometries, results without Coulomb potential are also
shown (labeled as Coul OFF). The results with Coulomb potential
are labeled as Coul ON. It is clear from the figure that the
effect of mass asymmetry on the $E_{bal}$ increases with increase
in impact parameter for all mass ranges. One can also see the
enhancement in $E_{bal}$ without Coulomb potential. This is caused
by the decrease in the repulsive interactions. In Fig. 1(c), data
points of $^{40}Ca+^{40}Ca$(${\eta}$=0 and $A_{TOT}$ = 80) and
$^{40}Ar+^{45}Sc$(${\eta}$=0.06 and $A_{TOT}$ = 85) are also
displayed \cite{12}. The match with data is not seen because the
SMD equation of state is used in the present study and $E_{bal}$
is found to be highly sensitive toward the EOS and the nn cross
section \cite{1,2,3}. But the trend of variation with impact
parameter is found to be the same.

In Fig. 2, we display the percentage change in balance energy
$\Delta E_{bal}^{b/b_{max}}$(\%), defined as $\Delta
E_{bal}^{b/b_{max}}$(\%) =
(($E_{bal}^{b/b_{max}\neq0.25}$-$E_{bal}^{b/b_{max}=0.25}$)/$E_{bal}^{b/b_{max}=0.25}$)$\times$100
versus the asymmetry of the reaction. Lines represents the mean
value of variation. Very interestingly, we see that for every
fixed $A_{TOT}$, the effect of impact parameter variation is
almost uniform throughout the asymmetry range. The values of mean
variation are given in Table I. It is very clear from Table I
that, that without Coulomb potential, the mean variation is
greater compared to that with Coulomb potential. For the symmetric
systems, it was found that the effect of the impact parameter on
$E_{bal}$ is greater for the lighter masses than for the heavier
ones \cite{11a}. A similar trend was observed in the present
study. We found that the mean variation decreases with increase in
$A_{TOT}$. In total, it is clear from Fig. 2, that the effect of
impact parameter variation is independent of ${\eta}$.
\begin{table*}
\caption{The values of ${\tau^{0.25}}$, ${\tau^{0.5}}$, and
${\tau^{0.75}}$ for ${\eta}$ = 0.1, 0.3, 0.5, and 0.7 for
calculations with and without Coulomb potential.}
\begin{center}
\begin{tabular}{|c|c|c|c|c|c|c|}
\hline ${\eta}$
&\multicolumn{2}{c|}{${\tau^{0.25}}$}&\multicolumn{2}{c|}{${\tau^{0.5}}$}&\multicolumn{2}{c|}{${\tau^{0.75}}$}\\\cline{2-7}

 &Coul. ON &Coul. OFF &Coul. ON &Coul. OFF &Coul. ON &Coul. OFF\\\cline{2-7}
 \hline 0.1  &-0.30  &-0.26 &-0.34 &-0.28 &-0.52 &-0.42 \\\cline{2-7}

 \hline 0.3  &-0.30  &-0.27 &-0.36 &-0.31 &-0.53 &-0.41 \\\cline{2-7}

 \hline 0.5  &-0.29  &-0.27 &-0.35 &-0.31 &-0.50 &-0.45 \\\cline{2-7}

 \hline 0.7  &-0.28  &-0.25 &-0.36 &-0.33 &-0.50 &-0.45 \\\cline{2-7}

 \hline
\end{tabular}
\end{center}
\end{table*}
In Fig. 3, we display the percentage difference $\Delta
E_{bal}^{\eta}$(\%) defined as $\Delta E_{bal}^{\eta}$(\%) =
(($E_{bal}^{\eta\neq0}$ -
$E_{bal}^{\eta=0}$)/$E_{bal}^{\eta=0}$)$\times$100 versus the
reduced impact parameter ($b/b_{max}$). Lines are the linear fits
($\propto m\frac {b}{b_{max}}$). We see that the effect of the
asymmetry variation increases with increase in the impact
parameter for each mass range. It means that for peripheral
collisions, the role of mass asymmetry is more as compared to
central ones. This is due to the fact that with increase in impact
parameter, the number of nn binary collisions decreases and the
increase of mass asymmetry further adds the same effect.

In Fig. 4, we display $E_{bal}$ as a function of $A_{TOT}$ for
$b/b_{max}$ = 0.25, 0.5, and 0.75, keeping ${\eta}$ fixed as 0.1
[Figs. 4(a) and 4(b)], 0.3 [Figs. 4(c) and 4(d)], 0.5 [Figs. 4(e)
and 4(f)], and 0.7 [Figs. 4(g) and 4(h)]. The results without
Coulomb potential are also shown. Various symbols are explained in
the caption of the figure. The lines are power-law fits ($\propto
A^{\tau}$). From the figure, we see that $E_{bal}$ follows the
power-law behavior at all values of ${\eta}$ and colliding
geometry. The values of ${\tau}$ are given in Table II.

From Fig. 4, we see that for each value of ${\eta}$, as we go from
central collisions to peripheral ones, $E_{bal}$ increases. The
increase is more for the lighter systems as compared to heavier
ones. From the values of ${\tau}$ in Table II, we see that as we
go from central to peripheral collisions, the value increases
drastically for each ${\eta}$. For the symmetric systems, the
values of ${\tau}$ are -0.32, -0.34, and -0.36, respectively, at
$b/b_{max}$ = 0.3, 0.45, and 0.6 for the $HMD^{40}$ (momentum
dependent hard equation of state with constant cross section of
40mb) EOS \cite{12a}. Our results are for the SMD EOS and at
$b/b_{max}$ = 0.25, 0.5, and 0.75, but the trend is still the
same. The values of ${\tau}$ without Coulomb potential are less
than the values with Coulomb potential, but the trend is the same.
The variation of ${\tau}$ as a function of $b/b_{max}$ and
${\eta}$ is shown in Fig. 5.

In Fig. 5, we display the variation of ${\tau}$ as a function of
$b/b_{max}$ for ${\eta}$ = 0.1-0.7 [Figs. 5(a) and 5(c)] and
variation with respect to ${\eta}$ for $b/b_{max}$ = 0.25-0.75
[Figs. 5(b) and 5(d)]. Dotted lines in [Figs. 5(b) and 5(d)]
represent the values of ${\tau}$ averaged over ${\eta}$. Symbols
are explained in the caption of the figure. From the figure, it is
clear that ${\tau}$ increases with increase in impact parameter
both with and without Coulomb potential. Very interestingly, the
variation of ${\tau}$ with ${\eta}$ is almost uniform as one goes
from central to peripheral collisions. It means that the mass
dependence of $E_{bal}$ is independent of mass asymmetry at every
colliding geometry. The values of ${\tau}$ averaged over ${\eta}$
are 0.29 (0.26), 0.35 (0.30), and 0.51 (0.43), respectively, for
$b/b_{max}$ = 0.25, 0.5, and 0.75 with (without) Coulomb
potential. The effect of Coulomb potential is greater for the
heavier nuclei; therefore, the increase in $E_{bal}$ is greater in
the case of heavier nuclei as compared to lighter ones when the
Coulomb potential is turned off. This leads to a decrease in value
of ${\tau}$ when calculations are performed without Coulomb
potential.

In summary, we studied the role of colliding geometry in the
disappearance of flow as well as its mass dependence throughout
the mass range 40-240 for mass-asymmetric reactions with ${\eta}$
= 0.1-0.7. Our results clearly demonstrate that the effect of mass
asymmetry is more dominant at peripheral collisions than at
central ones. Very interestingly, we found that the percentage
change in $E_{bal}$ with colliding geometry remains uniform
throughout the mass-asymmetry range for every fixed system mass.

Author is thankful to Dr. Rajeev K. Puri for interesting and
constructive discussions. This work is supported by a research
grant from the Council of Scientific and Industrial Research
(CSIR), government of India, vide Grant No.
09/135(0563)/2009-EMR-1.


\end{document}